\documentclass[preprint,aps,showpacs]{revtex4}
\def\ee{\end{equation}}
\def\be{\begin{equation}}
\def\ba{\begin{eqnarray}}
\def\ea{\end{eqnarray}}
\def\ll{\label}

\begin{document}

\title{Mutual thermal equilibrium in long-range Ising model using nonadditive
entropy}
\author{Ramandeep S. Johal}   
\email{rrai@lycos.com}   
 \affiliation{Post-graduate Department of Physics, Lyallpur Khalsa College, 
 Jalandhar-144001, India.}
\date{\today}

\begin{abstract}
Mutual equilibrium in long-range interacting systems
which involve nonadditive energy, is effectively described in terms of entropy with a nonadditive composition rule. As an example, long range Ising model is 
considered. The generality of the term having product of the system entropies  
is pointed out in this framework. 
\end{abstract}


\maketitle

\section*{1. Introduction}
Additivity of quantities such as entropy and energy, is an essential 
premise of thermodynamics and statistical mechanics of systems with short 
range interactions \cite{Callen1985}. Arguments about positivity of specific 
heat or concavity of the entropy-energy curve are based on this additivity 
postulate. On the other hand, systems where long range interactions are 
relevant, are basically nonadditive. Due to this feature, thermodynamics of 
such systems displays unusual properties, like inequivalence of different 
ensembles, negative specific heat in microcanonical ensemble and possible 
temperature discontinuity at first order transitions \cite{Dauxois2002}.
 For nonadditive systems, the non-concave entropy may be the correct entropy 
 of the system, with the result, the specific heat may be negative 
 \cite{Antonov1962, Lynden1968, Lynden1999}.

Similarly, the notion of macroscopic thermal equilibrium, which introduces the concept of temperature through the zeroth law, is based on additivity of entropy and energy. On the other hand, for large systems with long range interactions or even finite systems with short range interactions, the concept of equal temperatures over subsystems, does not hold at thermal equilibrium (maximum of total entropy). Clearly, if additivity postulate is relaxed for  either entropy or energy, we have breakdown of the zeroth law. 

In this paper, we consider two macroscopic systems in thermal contact with
each other, whose particles experience long range interactions. The total system
is isolated and so its energy is fixed. We consider the problem of
maximization of the total entropy under the constraint of fixed total energy.
Two alternate approaches are explained: the first one is based on the locality
and local extensivity assumptions \cite{Oppenheim2003}. Here all the correlations due to interactions are completely described by 
correlations in the energies of the systems. The total energy  $E$ is nonadditive and in general given by
\be
E = E_1 + E_2 + G(E_1, E_2),
\ll{none}
\ee
where  $E_1$ and $E_2$ are the energies of  system 1 and system 2 respectively, in the absence of interactions $G(E_1, E_2)$. These are called  local energies. Note that these are not the current energies of the subsystems which become modified due to the presence of interactions. The total entropy of the two systems, written in terms of $E_1$ and $E_2$, is additive:
\be
S(E_1,E_2) = S_1 (E_1) + S_2 (E_2).
	\ll{entsum}
\ee
Now upon maximization of the total entropy under constraint of Eq. (\ref{none}), 
one obtains that temperatures of the
two systems are not equal and the procedure fixes only the ratio of the two 
temperatures, unless entropy is defined by, say Boltzmann's principle. 

In this paper, we also study the approach, complementary to the above, where all
correlations due to interactions are absorbed in the entropy. The entropy so defined will naturally obey a generalized (nonadditive) composition rule. On the other hand, the energies in this alternate discription will be taken as additive. Thus our motivation is to see how nonadditive energies are effectively described in terms of the nonadditive entropy and what that entropic rule actually looks like.

The paper is arranged as follows: In section 2A, we treat the case when  Eqs. (\ref{none}) and (\ref{entsum}) are assumed; section 2B presents the alternate approach whereby energy is assumed to be additive. Thus we introduce the properties of generalized entropy and do the maximization with an additive
energy constraint.
 In section 3A, we take a specific long-range interacting model, namely
a spin-lattice model, and describe the maximization of entropy as in the approach of section 2A mentioned above. The final section 4 is devoted to  conclusions.
\section*{2A. Generic nonadditivity of energy}
Let us consider the variation of the total energy $E = E(E_1, E_2)$,
\be
dE = \left( \frac{\partial E}{\partial E_1} \right)_{E_2} dE_1
     + \left( \frac{\partial E}{\partial E_2} \right)_{E_1} dE_2.
     \ll{vare}
\ee
From Eq. (\ref{none}), we obtain
\be
\left( \frac{\partial E}{\partial E_1} \right)_{E_2} = 
1 + \left( \frac{\partial G}{\partial E_1} \right)_{E_2} \equiv \alpha_1 ,
	\ll{pde1}
\ee
and
\be
\left( \frac{\partial E}{\partial E_2} \right)_{E_1} = 
1 + \left( \frac{\partial G}{\partial E_2} \right)_{E_1} \equiv \alpha_2.
	\ll{pde2}
\ee
Due to constraint of the fixed total energy, we obtain
\be
dE = 0 \to dE_1  = - \frac{\alpha_2}{\alpha_1} dE_2.
	\ll{vareo}
\ee
Now if the total entropy is given by eq. (\ref{entsum}), we also impose 
$dS=0$, which implies
\be
\frac{ dS_1}{dE_1} dE_1  + \frac{ dS_2}{dE_2} dE_2   =0,
	\ll{vars}
\ee
or  
\be
\alpha_2 \frac{ dS_1}{dE_1}  = \alpha_1 \frac{ dS_2}{dE_2}.
	\ll{alps}
\ee
Here $\frac{ dS_i}{dE_i} = \frac{1}{T_i}$ is defined to be the inverse local temperature of each system \cite{Oppenheim2003}. Clearly, these temperatures
 are not equal in the presence oflong range interactions. 

\section*{2B. Second approach: Nonadditive entropy but additive energy}
Now we formulate an equivalent description in terms of nonadditive
entropy. We assume that all the previous correlations expressed through 
energy are absorbed in the generalized entropy. Thus the energy in our 
alternate description is taken as additive. Our goal is to seek the 
composition rule for entropy, so that upon maximization of total entropy 
under the constraint of fixed total energy, we obtain a form similar to Eq. (\ref{alps}).
   
We suppose that the generalized entropy 
of the composite system made up of two systems $1$ and $2$, can be written as 
\be
{S}(1,2) = f({S}_1,{S}_2),
\ll{compose}
\ee
where ${S}_1(E_1)$ and ${S}_2(E_2)$ are the system entropies, as functions of
respective local energies. Note that although we use here the same symbols for
entropies and energies as in the previous section, these two sets, clearly are 
different quantities. $f$ is a certain bivariate function, such that the function itself and all its derivatives are continuous. 

Thus consider the sitation when the total energy of the system is fixed 
at ${E}$. Let ${E}_1$ and  ${E}_2$ be the energies of the systems when the 
total entropy $f({S}_1, {S}_2)$ is maximized. Thus putting
$df = 0 $, under the constraint of additive total energy;
\be
{E} = {E}_1 +  {E}_2,
\ll{sume}
\ee
we obtain 
\be
\left( \frac{ \partial f}{\partial {S}_1} \right)_{{S}_2} 
       \frac{ d{S}_1}{d {E}_1} d {E}_1 +  
    \left( \frac{ \partial f}{\partial {S}_2} \right)_{{S}_1}    
     \frac{ d{S}_2}{d {E}_2} d {E}_2   =0.
	\ll{varf0}
\ee
With $d{E}=0$, we have $d {E}_1 = - d {E}_2$. Thus Eq. (\ref{varf0})
can be written as
\be
\frac{ \partial f}{\partial {S}_1} \frac{ d{S}_1}{d {E}_1}
= \frac{ \partial f}{\partial {S}_2} \frac{ d{S}_2}{d {E}_2}. 
	\ll{fs}
\ee
Now when we seek a description equivalent to the first approach, 
 we require that Eq. (\ref{fs}) be of same form as  Eq. (\ref{alps}).
Thus the following conditions hold
\ba
\frac{\partial f}{\partial {S}_1} & = &
\phi({E}_1, {E}_2) \left[ 1 +  \frac{\partial G}{\partial E_2} \right]  
\ll{fs1g} \\
\frac{\partial f}{\partial {S}_2} & = &
\phi({E}_1, {E}_2) \left[ 1 + \frac{\partial G}{\partial E_1}  \right],
 \ll{fs2g}
\ea
where $\phi$ is some unknown differentiable function.
Here it is apparent that the composition rule for the generalized 
entropy will turn out to be ambiguous due to the arbitrary function $\phi$. However, it is
possible to make a specific choice, as following. Note from Eq. (\ref{none}) that
when $G=0$, the total energy is additive over local energies
of the systems. In this situation, it is expected that 
$\frac{\partial f}{\partial {S}_i} =1$, i.e. entropy is additive. Thus in the following, 
we set the function $\phi$ as a constant equal to unity.

Now we evaluate the following cross derivatives
\ba
\frac{\partial^2 f}{\partial {S}_2  \partial {S}_1} &=& 
\frac{\partial^2 G}{\partial {S}_2  \partial {E}_2} = 
\frac{ dE_2}{dS_2} \frac{\partial^2 G}{\partial {{E}_2}^2} \ll{fs21} \\
\frac{\partial^2 f}{\partial {S}_1  \partial {S}_2} &=& 
\frac{\partial^2 G}{\partial {S}_1  \partial {E}_1} = 
\frac{ dE_1}{dS_1} \frac{\partial^2 G}{\partial {{E}_1}^2} \ll{fs12}.
\ea
Assuming that the function $f$ and its derivatives are continuous, 
we get the equality of the above cross derivatives, yielding
\be
\frac{ dE_1}{dS_1} \frac{\partial^2 G}{\partial {{E}_1}^2} =
\frac{ dE_2}{dS_2} \frac{\partial^2 G}{\partial {{E}_2}^2}.
\ll{eqcross}
\ee
We simplify the above condition by invoking separation of variables. For this 
we assume that 
$\frac{\partial^2 G}{\partial {{E}_i}^2}$, involve
variables or parameters of the respective system ($i$) only.
The simplest case is when 
\be
\frac{\partial^2 G}{\partial {{E}_i}^2} = \nu_i,
\ll{doublege1}
\ee
where $\nu_i$ is constant and may depend upon the pre-specified parameters of the system. Thus we have
\be
\frac{ dE_1}{dS_1} \nu_1 =
\frac{ dE_2}{dS_2} \nu_2 = \chi,
\ll{sep1}
\ee
where $\chi$ is a constant.
These equations can be integrated to give
\be
 {S}_i = \frac{\nu_i}{\chi} {E}_i + \sigma_i,
 \ll{expst}
\ee
where $(i=1,2)$ and $\sigma_i$ is  constant of integration. 

Moreover, $G$ of Eq. (\ref{none}) is a known function of $E_1$ and $E_2$, so that
\be
\frac{\partial G}{\partial {{E}_i}} = \nu_i E_i + \psi(E_j)
\ll{singlege1}
\ee
where $i\ne j$ and $\psi$ is a known function.

Eq. (\ref{singlege1}) can be expressed in terms of $S_{i(j)}$ using (\ref{expst}).
Then Eqs. (\ref{fs1g}) and (\ref{fs2g}) (with $\phi =1$) can be integrated to obtain the generalised
composition rule for the entropy.
\section*{3A. Long range Ising model: an example}
In this section, we take a specific long range interacting model as an example of the general treatment presented so far.
Consider a lattice  of $N$ spins with total energy as given by

\be
E = h_1 e_1  + h_2 e_2 - \frac{J_1}{2} {e_1}^2 - \frac{J_2}{2} {e_2}^2 
	- J_{12} e_1 e_2.
	\ll{hamilse}
	\ee
The total system is assumed to be separated into two regions with a fixed 
number of sites.
The spin excess $e_i = N_{i}^{+}  - N_{i}^{-}$, where $N_{i}^{\pm}$ are the number of up (+) and down (-) spins inside each region. 
$J_i$ and $h_i$ are the known coupling constants within each region.
$J_{12}$ represents the coupling between each region, assumed to be smaller
than $J_i$. Note that the local energy of each region is $E_i = h_i e_i$.
%
%
%
Writing $E$ in terms of $E_1$ and $E_2$, 
consider the variation of the total energy, $dE(E_1, E_2)$ and 
note that
\be
\left( \frac{\partial E}{\partial E_1} \right)_{E_2} = 
1 - \frac{J_1}{{h_1}^2 } {E_1} - \frac{J_{12}}{h_1 h_2} E_2 \equiv \alpha_1,
	\ll{pde1ii}
\ee
and
\be
\left( \frac{\partial E}{\partial E_2} \right)_{E_1} = 
1 - \frac{J_2}{{h_2}^2 } {E_2} - \frac{J_{12}}{h_1 h_2} E_1 \equiv \alpha_2.
	\ll{pde2ii}
\ee
Imposing the maximisation of total entropy (additive) under the  constraint 
of a fixed total energy $E$, as discussed in  section 2A,  
the equilibrium condition, Eq. (\ref{alps}) takes the following form
\be
\left[1 - \frac{J_2}{{h_2}^2 } {E_2} - \frac{J_{12}}{h_1 h_2} E_1 \right] \frac{ dS_1}{dE_1}  = 
 \left[1 - \frac{J_1}{{h_1}^2 } {E_1} - \frac{J_{12}}{h_1 h_2} E_2\right] \frac{ dS_2}{dE_2}.
	\ll{alps2}
\ee
\section*{3B. Long range Ising model: second approach}
Following the general procedure, described in the beginning of section 2B, we now demand that applying the second approach to this model we obtain  the same equilibrium condition as (\ref{alps2}). Thus the following conditions hold 
\ba
\frac{\partial f}{\partial {S}_1} & = &
\phi({E}_1, {E}_2) \left[ 1 - \frac{J_2}{{h_2}^2 } {{E}_2} 
- \frac{J_{12}}{h_1 h_2} {E}_1 \right], 
\ll{fsg1}   \\
\frac{\partial f}{\partial {S}_2} & = & 
\phi({E}_1, {E}_2) \left[1 - \frac{J_1}{{h_1}^2 } {{E}_1} 
- \frac{J_{12}}{h_1 h_2} {E}_2 \right],
\ll{fsg2}
\ea
Here also, as in section 2B, we restrict to the case when $\phi =1$.

Then equality of the cross derivatives of $f$ yields the following condition
\be
\frac{J_1}{{h_1}^2} \frac{d {E}_1}{d {S}_1} =
\frac{J_2}{{h_2}^2} \frac{d {E}_2}{d {S}_2}.
\ll{eq12}
\ee
Thus the constant $\nu_i$ of Eq. (\ref{sep1}) is expressed in terms of the parameters of the individual systems. Now, 
as each side pertains to an individual sytem, so we equate each side to a constant $\chi$. Finally, we obtain
\be
\frac{d {S}_i}{d {E}_i} = \frac{J_i}{\chi {h_i}^2}, \quad i =1,2,
\ll{tempt}
\ee
which effectively implies a linear relation between the entropy of the system
and its energy
\be
 {S}_i = \frac{J_i}{\chi {h_i}^2} {E}_i + \sigma_i.
 \ll{expstii}
\ee
Here $\sigma_i$ is  constant of integration. It can be determined from
the fact that $S_i = 0$ for $E_i = -h_i N_i$. Thus $\sigma_i = \frac{J_i N_i}
{\chi h_i}$.
 
Using the above relation, we can write Eq. (\ref{fsg1}) (with $\phi =1$) as follows
\be
\frac{ \partial f}{\partial {S}_1} =
\left[ 1 - \chi ({S}_2 - \sigma_2 ) - \chi \frac{J_{12} h_1}{ J_1 h_2} 
({{S}_1} - \sigma_1) \right],
\ll{fsfin}
\ee
and a similar relation corresponding to Eq. (\ref{fsg2}).
These expressions can be easily integrated to obtain the form of
total entropy in terms of the system entropies
\be
f({S}_1, {S}_2 ) =
c_1 {S}_1 + c_2 {S}_2 - 
\frac{\chi J_{12} h_1 }{2 J_1 h_2} {{S}_1}^2 -  \frac{\chi J_{12} h_2 }{2 J_2 h_1}  {{S}_2}^2
- 2 \chi {S}_1 {S}_2.
\ll{fsol}
\ee
This is a form of nonadditive entropy rule that effectively describes
the mutual equilibrium between two long range Ising model systems.
The constants $c_i$ are explicitly given by
\ba
c_1 &=& 1 + \frac{J_2 N_2}{h_2} + \frac{J_{12} N_1 }{h_2}, \\
c_2 & = & 1 + \frac{J_1 N_1}{h_1} + \frac{J_{12} N_2 }{h_1}.
\ea 
\section*{4. Conclusions}
We have mapped the mutual thermal equilibrium of a bipartite long range interacting spin model, in terms of additive total entropy and
nonadditive total energy, to a description in terms of nonadditive total 
entropy and additive total energy.
The composition rule for the nonadditive entropy has been derived. 
Additionally, we make the following observations:
for $J_{12} =0$, the total entropy is in the form
$f = c_1 S_1 + c_2 S_2 - 2 \chi S_1 S_2$. The additional product term is due to the fact that
there are still interactions within each system, represented by the coupling constants $J_1$ and $J_2$.  

We comment on the generality of this product term, which means it always 
appears in the entropy composition rule, when in Eqs. (\ref{eqcross}) 
we assume that
\be
\frac{\partial^2 G}{\partial {{E}_i}^2} = g(E_i),
\ll{doublege2}
\ee
where $g$ is some integrable function of its argument. 

To show this, note that due to separation of variables, we can write
\be
\frac{ dE_1}{dS_1} g(E_1) =
\frac{ dE_2}{dS_2} g(E_2) = \xi,
\ll{sep1p}
\ee
where $\xi$ is a constant of separation. These equations can be integrated to give
\be
 \xi {S}_i =  \int g(E_i) \; d{E}_i + const.
 \ll{expst2}
\ee
Integrating Eq. (\ref{doublege2}) and using Eq. (\ref{expst2})
in the result so obtained, we can write
\be
\frac{\partial G}{\partial {{E}_i}} = \xi S_i - const + \psi(E_j)
\ll{singlege1p}
\ee
where $i\ne j$ and $\psi$ is a known function.
Again, using the above Eq. in Eqs. (\ref{fs1g}) and (\ref{fs2g}), we see that
the obtained $f$ function after integration, necessarily contains the term $\xi S_1 S_2$. 
\section*{Acknowledgements}
It is a pleasure to thank the organisers of the NEXT-2005 conference,
Giorgio Kaniadakis and Anna Carbone, and Stefano Ruffo for giving me
opportunity to attend this conference. 

\end{document}